\documentclass[aps,twocolumn,showpacs,preprintnumbers,prb]{revtex4}
\usepackage{graphicx}
\usepackage{amssymb}

\newcommand{\R}{\mathbf{r}}

\newcommand{\UP}{n_{\uparrow}}
\newcommand{\DN}{n_{\downarrow}}

\newcommand{\be}{\begin{equation}}
\newcommand{\ee}{\end{equation}}
\newcommand{\bea}{\begin{eqnarray}}
\newcommand{\eea}{\end{eqnarray}}
\newcommand{\bean}{\begin{eqnarray*}}
\newcommand{\eean}{\end{eqnarray*}}

\begin{document}

\title{Exchange-Correlation Hole of a Generalized Gradient Approximation \\
for Solids and Surfaces}
\author{Lucian A. Constantin$^1$, John P. Perdew$^1$, and J. M.
Pitarke$^{2,3}$}
\affiliation{
$^1$Department of Physics and
Quantum Theory Group, Tulane University, New Orleans, LA 70118\\
$^2$CIC nanoGUNE Consolider, Mikeletegi Pasealekua 56, E-20009 Donostia, Basque
Country\\
$^3$Materia Kondentsatuaren Fisika Saila (UPV/EHU), DIPC, and Centro F\'\i
sica Materiales CSIC-UPV/EHU,\\
644 Posta kutxatila, E-48080 Bilbo, Basque Country}

\date{\today}

\begin{abstract}
We propose a generalized gradient approximation (GGA) for the angle- and 
system-averaged  
exchange-correlation hole of a many-electron system. This hole, which 
satisfies known exact 
constraints, recovers the PBEsol (Perdew-Burke-Ernzerhof for solids) exchange-correlation 
energy functional, a GGA that accurately describes the equilibrium 
properties of densely packed solids and their surfaces. 
We find that our PBEsol exchange-correlation hole describes 
the wavevector analysis of the jellium exchange-correlation 
surface energy in agreement with a sophisticated
time-dependent density-functional calculation (whose three-dimensional wavevector
analysis we report here). 
\end{abstract}

\pacs{71.10.Ca,71.15.Mb,71.45.Gm}

\maketitle

\section{Introduction}
\label{sec1}

In the Kohn-Sham (KS) density functional theory \cite{KS} for 
the ground-state energy of a many-electron system, only the 
exchange-correlation (xc) energy $E_{xc}[n]$ has to be approximated. 
The exact xc energy of an arbitrary inhomogeneous system of
density $n({\bf r})$, which incorporates all the quantum many-body 
effects beyond the Hartree approximation, can be obtained from the spherical average
$\bar n_{xc}({\bf r},u)$ of the coupling-constant averaged xc hole density
$\bar n_{xc}({\bf r},{\bf r}')$ at ${\bf r}'$ around an electron at ${\bf r}$ as
follows \cite{LP1,PCP} 
\begin{equation}
E_{xc}[n]=\int d{\bf r}\,n({\bf r})\,\varepsilon_{xc}[n]({\bf r}),
\label{e1}
\end{equation}
where $\varepsilon_{xc}[n]({\bf r})$ is the xc energy per particle at point
${\bf r}$:
\begin{eqnarray}\label{new}
\lefteqn{
\varepsilon_{xc}[n]({\bf r})=\frac{1}{2}\int^\infty_0 du\; 4\pi
u^2\frac{1}{u}\bar n_{xc}(\R,u)}\nonumber\\
& =4\int^{\infty}_{0} dk\int^{\infty}_{0} du\,u^2\,
{\sin ku\over ku}\,\bar n_{xc}({\bf r},u),
\label{e2}
\end{eqnarray}
with
\begin{equation}
\bar n_{xc}({\bf r},u)=
{1\over 4\pi}\int d\Omega\,\bar n_{xc}({\bf r},{\bf r}'),
\label{e3}
\end{equation}
$d\Omega$ being a differential solid angle around the direction of
${\bf u}={\bf r}'-{\bf r}$, and $k$ representing the magnitude of the wavevector. 
(Unless otherwise stated, atomic units are used throughout,
i.e., $e^2=\hbar=m_e=1$.)

The "Jacob's ladder" classification of the widely-used ground-state
density-functional
approximations for $E_{xc}[n]$ and $\bar n_{xc}({\bf r},u)$ has three complete 
non-empirical rungs: the local-spin-density approximation (LSDA),~\cite{KS} 
the generalized-gradient approximation (GGA),~\cite{PBE,PBW,EP} and 
the meta-GGA.~\cite{TPSS,CPT} Due to its simplicity and accuracy, 
one of the most commonly used xc density-functional approximation in solid-state
physics 
and quantum chemistry calculations is nowadays the semilocal PBE-GGA.\cite{PBE}

Recent work\cite{PCSB} has shown, however, that 
the exchange 
density-functional approximations should recover, in the limit of 
slowly-varying densities, the universal second-order gradient-expansion (GE) 
approximation of the exchange energy, 
\begin{equation}
E^{GE}_x[n]=\int d\R\;n(\R)\epsilon^{unif}_x(n(\R))[1+\mu^{GE}_x s^2(\R)+...],
\label{e4b}
\end{equation}
where $\epsilon^{unif}_x$ is the exchange energy per particle of the uniform electron 
gas, $\mu^{GE}_x=10/81$ is the GE exchange coefficient,\cite{AK} and
$s=|\nabla n|/(2k_Fn)$ is the
reduced density gradient which measures the variation of the electron 
density over a Fermi wavelength $\lambda_F=2\pi/k_F$, 
with $k_F=(3\pi^2 n)^{1/3}$ representing the magnitude of the local 
Fermi wavevector. 
Recovery of the correct second-order
gradient expansion for correlation \cite{MB} in the slowly-varying limit is 
much less important for the
construction
of density-functional approximations.
(See Table 1 of Ref.~\onlinecite{PCSB}).

A GGA, which has 
as ingredients only the spin densities $\UP$ and $\DN$ and 
their gradients $\nabla \UP$ and $\nabla\DN$, cannot recover, in the 
limit of slowly varying densities, the GE approximation of the exchange 
energy and at the same time be accurate for atoms.\cite{PCSB,PRCVSCZB} 
The semilocal PBE has the 
correct correlation GE coefficient in the high-density limit,
and is accurate for atoms, 
but its exchange GE coefficient is almost twice as large 
as the exact coefficient, i.e.,
$\mu_x^{PBE}\approx 2\mu_x^{GE}$. Because of this,\cite{PRCVSCZB} PBE 
overestimates the equilibrium lattice constants of solids and 
yields surface energies that are too low.

Following the ideas of Ref.~\onlinecite{PCSB}, PBEsol (PBE for solids) 
was constructed \cite{PRCVSCZB}. PBEsol is a GGA that has the same form as 
PBE but restores the density-gradient expansion for exchange by 
replacing $\mu^{PBE}_x=0.2195$ with 
$\mu^{PBEsol}_x=\mu^{GE}_x$. By fitting the jellium xc surface 
energies (as had been done previously in Ref.~\onlinecite{AM05} 
in the construction of a GGA which relies on the Airy-gas 
approximation\cite{KM}), the PBEsol correlation GE coefficient 
was set to $\mu^{PBEsol}_c=0.046$.
(For PBE, $\mu^{PBE}_c=0.0667$). Thus, PBEsol can easily be  
applied in solid-state calculations 
(just by changing the coefficients in a PBE code)
and yields good equilibrium 
lattice constants and jellium surface energies.\cite{PRCVSCZB} 
Several other applications of PBEsol
have already proved the 
accuracy of this GGA for solids. In particular, PBEsol 
considerably improves the structure of gold clusters \cite{Furche1} 
and works better than PBE for isomerization energies and 
isodesmic stabilization energies of hydrocarbon molecules.\cite{Gabor}       
PBEsol also describes ferro- and anti-ferro-electric $ABO_3$ crystals \cite{WVK} much
better than LSDA or PBE-GGA.

In this paper, we first construct the PBEsol angle-averaged xc hole density
$\bar n^{PBEsol}_{xc}({\bf r},u)$. A nonempirical derivation of the 
PBE xc hole was reported in Ref.~\onlinecite{PBW}, starting from 
the second-order density-gradient expansion of the xc hole 
and cutting off the spurious large-$u$ contributions to 
satisfy exact constraints according to which (i) the 
exchange-hole density must be negative, (ii) the exchange 
hole must integrate to -1, and (iii) the correlation hole 
must integrate to zero. Later on, a fully smoothed analytic 
model was constructed for the PBE exchange hole.\cite{EP} 
Our construction of the PBEsol xc hole 
begins with and
appropriately modifies the sharp
cutoff {\it correlation} hole of Ref.~\onlinecite{PBW}   
and the smooth {\it exchange} hole of Ref.~\onlinecite{EP}.
It should be recalled that, because of an integration by parts that occurs
in the underlying gradient expansion, a GGA hole is meaningful only after averaging
over the electron density $n(\R)$ (as in our tests and applications),
and this system-averaging itself smooths sharp cutoffs.

Finally, we use our PBEsol xc hole to carry out a 
three-dimensional (3D) wavevector analysis of the 
jellium xc surface energy. This wavevector analysis 
was carried out in Ref.~\onlinecite{PCP} in the 
random-phase approximation (RPA). Here, we go beyond the 
RPA in the framework of time-dependent density-functional 
theory (TDDFT), and we compare these calculations with 
the results we obtain from our PBEsol xc hole density.

The paper is organized as follows. In Sec. II, 
we present the PBEsol angle-averaged xc-hole density 
$\bar n^{PBEsol}_{xc}({\bf r},u)$. In section III, 
we perform the wavevector analysis of the jellium 
xc surface energy. In Sec. IV, we summarize our conclusions. 

\section{PBEsol-GGA angle-averaged exchange-correlation hole}
\label{sec2}

We assume here that the PBEsol correlation energy can be constructed from a
gradient expansion for the correlation hole in much the same way that the PBE
correlation energy was so constructed \cite{PBW}.
The GGA angle-averaged correlation hole is \cite{PBW}
\begin{equation}
\bar n^{GGA}_c(r_s,\zeta,t,v)=\phi^5k^2_s[A_c(r_s,\zeta,v)+t^2B_c(r_s,\zeta,v)]\theta 
(v_c-v),
\label{e5}
\end{equation}
where $r_s = (9\pi/4)^{1/3}/k_F$ is a local density parameter, 
$\zeta=(\UP-\DN)/(\UP+\DN)$ is the relative spin polarization, 
$\phi=[(1+\zeta)^{2/3}+(1-\zeta)^{2/3}]/2$ is a spin-scaling factor, $v=\phi k_su$ 
with $k_s = (4 k_F/\pi)^{1/2}$
is the reduced electron-electron separation on the scale of the screening length, and 
$t=|\nabla n|/(2\phi k_sn)$ is the reduced density
gradient measuring the variation of the electron density over the
screening length. The 
sharp cutoff $v_c$ is found such that Eq.~(\ref{e5}) satisfies the correlation 
hole sum rule $\int d\R \;n_c(\R,\R')=0$.  
$\phi^5k^2_sA_c(r_s,\zeta,v)$ is the LSDA correlation hole\cite{PW1} given by
Eq.~(45) of Ref.~\onlinecite{PBW}, and the gradient 
correction to the correlation hole is given by the following expression:\cite{PBW}
\begin{equation}
B_c(r_s,\zeta,v)=B^{LM}_c(v)[1-e^{-pv^2}]+\beta(r_s,\zeta)v^2e^{-pv^2},
\label{e6}
\end{equation}
where $B^{LM}_c(v)$ is the RPA nonoscillating long-range contribution given by
Eq.~(49) of Ref.~\onlinecite{PBW}, $p(r_s,\zeta)=\pi k_F(0.305-0.136\zeta^2)/4\phi^4$ 
measures where 
the short-range contribution vanishes, and 
\begin{equation}
\beta(r_s,\zeta)=\frac{2p^2}{3\pi^3}[\frac{\mu^{GGA}_c}{\mu_c^{PBE}}- 
E_1(12p)]
\label{e7}
\end{equation}
is constructed so that the second-order gradient expansion of the PBEsol
correlation energy is recovered.
Here $E_1(y)=ye^y\int^{\infty}_ydt\;e^{-t}/t$ is between 0 and 1, and
$\mu^{GGA}_c$ 
is the GGA gradient coefficient in the slowly-varying limit 
($\mu^{PBE}_c$ for PBE and $\mu^{PBEsol}_c$ for PBE-sol).

\begin{figure}
\includegraphics[width=\columnwidth]{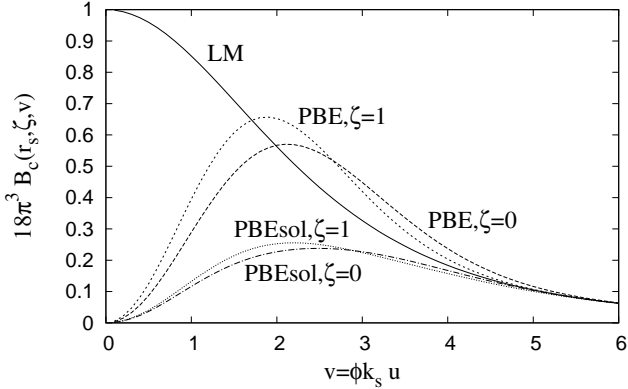}
\caption{Gradient correction to the correlation hole, $B_c(r_s,\zeta,v)$,
versus $v$ for $r_s=2$. 
PBE and PBEsol are compared here.
The solid line represents the Langreth-Mehl (LM) RPA 
contribution, which should be recovered at large $v$.}
\label{f1}
\end{figure}

In Fig.~\ref{f1}, we show PBE and PBEsol versions of $B_c(r_s,\zeta,v)$ 
versus $v$, for $r_s=2$. Both PBE and PBEsol recover 
the correct RPA-like behavior [$B^{LM}_c(v)$] at large $v$, 
and they both show the same $\zeta$ behavior; because 
$\mu^{PBEsol}_c<\mu^{PBE}_c$, however, at intermediate 
values of $v$ the PBEsol gradient correction to the 
correlation hole is substantially smaller than the PBE one. 
The gradient correction $B_c(r_s,\zeta,v)$ of Eq.~(\ref{e6}) 
can be negative at small $v$ and small $r_s$; however, 
because of the energy sum rule both
$\int^{\infty}_0 du\;u B_c(r_s,\zeta,v)$ and
$\int^{\infty}_0 du\;u^2 B_c(r_s,\zeta,v)$ are positive, 
which ensures that the cutoff procedure is correct and 
for every value of $r_s$, $\zeta$, and $t$ there is 
a $v_c$ such that Eq.~(\ref{e5}) satisfies the correlation-hole sum rule.
%
%

\begin{figure}
\includegraphics[width=\columnwidth]{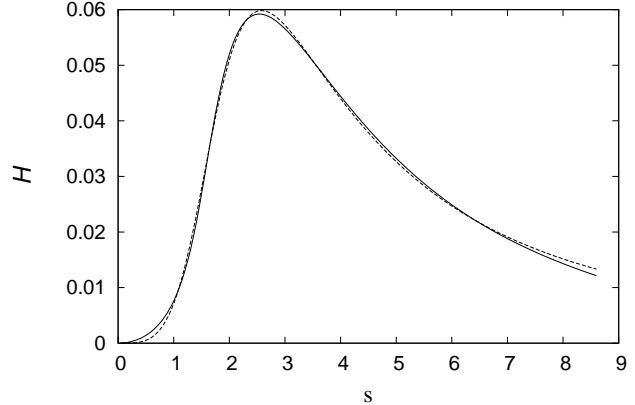}
\caption{The exponent $\mathcal{H}(s)$ of Eq.~(\ref{e10}) versus 
the reduced gradient $s$. The solid line represents the numerical 
solution of the implicit equation for $\mathcal{H}(s)$ 
[Eq.~(A4) of Ref.~\onlinecite{EP}]. The dashed line 
represents the fit of Eq.~(\ref{e14}).} 
\label{f2}
\end{figure}

The exchange energy and exchange-hole density for a spin-polarized system may be 
evaluated from their spin-unpolarized counterparts by 
using the spin-scaling relations\cite{PBW,OP}
\begin{equation}
E_{x}[n_{\uparrow} ,n_{\downarrow}]=\frac{1}{2}\{E_{x}[2 n_{\uparrow}]+E_{x}[2 
n_{\downarrow}]\}
\label{e8}
\end{equation}
and
\begin{equation}
n_{x}[n_{\uparrow} ,n_{\downarrow}](\mathbf{r},\mathbf{r}+\mathbf{u})=
\sum_{\sigma}\frac{n_{\sigma}(\mathbf{r})}{n(\mathbf{r})}n_{x}[2 n_{\sigma}]
(\mathbf{r},\mathbf{r}+\mathbf{u});
\label{e9}
\end{equation}
thus, we only need to consider the spin-unpolarized system.
As in the case of the analytical PBE exchange hole of Ref.~\onlinecite{EP}, 
we choose the following ansatz for the nonoscillatory dimensionless exchange-hole shape:
\begin{eqnarray}
\lefteqn{
J^{PBEsol}(s,y)=[-\frac{\mathcal{A}}{y^{2}}\frac{1}{1+(4/9)\mathcal{A}y^{2}}
 +(\frac{\mathcal{A}}{y^{2}}+\mathcal{B}+}\nonumber\\
&\mathcal{C}[1+s^2\mathcal{F}(s)]y^{2}+\mathcal{E}[1+s^2\mathcal{G}(s)]y^{4}
)e^{-\mathcal{D}y^{2}}]
e^{-s^{2}\mathcal{H}(s)y^{2}},
\label{e10}
\end{eqnarray}
where $s$ is the reduced density gradient for exchange. When $s=0$, 
Eq.~(\ref{e10}) recovers \cite{PW,EP} $J^{LSDA}$ for 
$\mathcal{A}=1.0161144$, 
$\mathcal{B}=-0.37170836$,
$\mathcal{C}=-0.077215461$, $\mathcal{D}=0.57786348$, 
and $\mathcal{E}=-0.051955731$. The functions 
$\mathcal{F}(s)$, $\mathcal{G}(s)$, and $\mathcal{H}(s)$ 
are found in such a way that the energy and exchange-hole sum rules are satisfied:
\begin{equation}
\frac{8}{9}\int^{\infty}_{0}dy\; y J^{PBEsol}(s,y)=-F^{PBEsol}_{x}(s)
\label{e11}
\end{equation}
and
\begin{equation}
\frac{4}{3\pi}\int^{\infty}_{0}dy\; y^{2}J^{PBEsol}(s,y)=-1,
\label{e12}
\end{equation}
and also the small-$u$ behavior of the exchange hole
is recovered by \cite{EP}:
\begin{equation}
\mathcal{F}(s)=6.475\mathcal{H}(s)+0.4797.
\label{e13}
\end{equation}
Here $F^{PBEsol}_{x}(s)$ is the PBEsol enhancement factor.\cite{PRCVSCZB}
The integrals of Eqs.~(\ref{e11}) and (\ref{e12}) can be solved analytically, 
\cite{EP} and Eqs.~(\ref{e11})-(\ref{e13}) reduce (by substitution) to an 
implicit equation for $\mathcal{H}$ [Eq.~(A4) of Ref.~\onlinecite{EP}].
We have solved this equation for PBEsol; the numerical solution that 
we have found for $\mathcal{H}(s)$ (see Fig.~\ref{f2}) can be fitted to the following analytic 
expression:
\begin{equation}
\mathcal{H}(s)=\frac{a_1s^2+a_2s^4}{1+a_3s^4+a_4s^6},
\label{e14}
\end{equation}
where $a_1=0.00018855 $, $a_2=0.00741358 $, $a_3=0.05687256$, and $a_4=0.00675093$.
For $s>8.5$, as occurs in the tail of an atom or molecule where the electron 
density is negligible, the implicit equation for $\mathcal{H}(s)$ 
does not have a solution (as in the PBE case \cite{EP}) so we reset $s$ to $s=8.5$.
Recently, Henderson \emph{et al.} \cite{HJS} constructed a GGA exchange hole that 
eliminates 
this unphysical large-$s$ behavior by using
some ideas from the meta-GGA hole \cite{CPT}.

\begin{figure}
\includegraphics[width=\columnwidth]{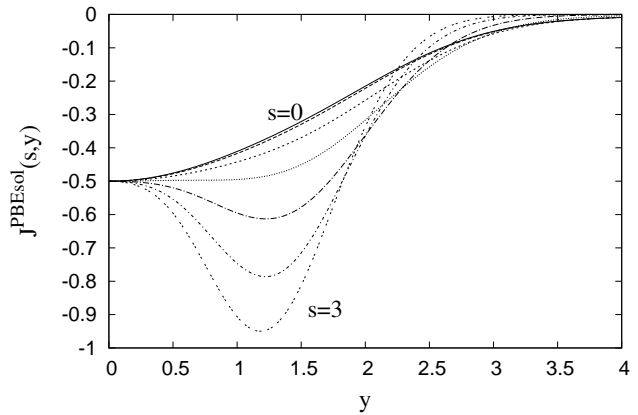}
\caption{Dimensionless exchange hole shape
$J^{PBEsol}(s,y)$ [see Eq.~(\ref{e10})] 
versus $y=k_{F}u$, for $s$ between 0 and 3 in steps of 0.5. For comparison with 
$J^{PBE}(s,y)$, see Fig.~2 of Ref.~\onlinecite{EP}. 
When $s=0$, $J^{PBEsol}(s,y)$ yields $J^{LSDA}(y)$.} 
\label{f3}
\end{figure}

In Fig.~\ref{f3}, 
we plot the dimensionless exchange hole shape
$J^{PBEsol}(s,y)$ 
[using the analytical fit of Eq.~(\ref{e14})] versus 
$y=k_Fu$ for several values of the reduced gradient $s$. 
Our $J^{PBEsol}(s,y)$ looks similar to the $J^{PBE}(s,y)$ 
of Ref.~\onlinecite{EP}, but $J^{PBE}(s,y)$ is deeper 
because $\mu^{PBE}_x=0.2195 > \mu^{PBEsol}_x=0.1235$.

\begin{figure}
\includegraphics[width=\columnwidth]{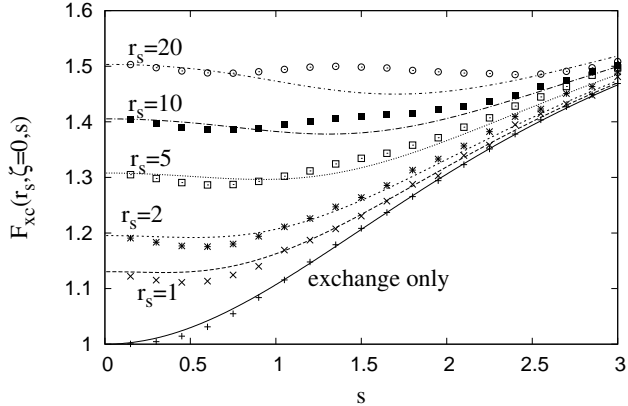}
\caption{The PBEsol enhancement factor $F_{xc}$ for the spin-unpolarized case
($\zeta=0$), as a function of the reduced gradient $s$ for several 
values of $r_s$. The lines represent the enhancement factor 
obtained from the PBEsol xc energy functional of 
Ref.~\onlinecite{PRCVSCZB}. The dots represent the 
enhancement factor obtained from our PBEsol angle-averaged xc hole density through Eq.~(\ref{e2}).}
\label{f4}
\end{figure}

\begin{figure}
\includegraphics[width=\columnwidth]{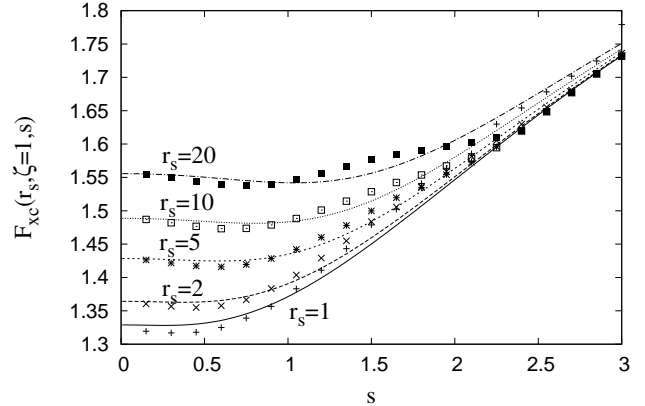}
\caption{The PBEsol enhancement factor $F_{xc}$ for the fully-spin-polarized case
($\zeta=1$), as a function of the reduced gradient $s$ for 
several values of $r_s$. The lines represent the enhancement 
factor obtained from the PBEsol xc energy functional of 
Ref.~\onlinecite{PRCVSCZB}. The dots represent the enhancement 
factor obtained from our PBEsol angle-averaged xc hole density through Eq.~(\ref{e2}).}
\label{f4b}
\end{figure}

Finally, we look at the xc enhancement factor, which displays the nonlocality:
\cite{PT}
\begin{equation}
F_{xc}^{GGA}=\frac{\epsilon_{xc}^{GGA}(\UP,\DN,\nabla\UP,\nabla\DN)}{\epsilon_x^{unif}(n)},
\label{e3b}
\end{equation}
$\epsilon_x^{unif}(n)$ being the exchange energy per particle of a spin-unpolarized 
uniform electron gas. 
For a spin-unpolarized system in the high-density limit ($r_s\rightarrow 0$) the exchange
energy is dominant and
Eq. (15) defines
the exchange enhancement factor $F^{GGA}_x=\epsilon^{GGA}_x(n,\nabla 
n)/\epsilon^{unif}_x(n)$.
In Figs. \ref{f4} and \ref{f4b}, 
we show the PBEsol enhancement factor for a spin-unpolarized system, 
$F^{PBEsol}_{xc}(r_s,\zeta=0,s)$, and for a fully-spin-polarized system, 
$F^{PBEsol}_{xc}(r_s,\zeta=1,s)$, versus $s$ for several values of $r_s$. 
$F^{PBEsol}_{xc}$ is calculated either (i) from the analytic 
expression of $\epsilon_{xc}^{PBEsol}$ reported in
Ref.~\onlinecite{PRCVSCZB} or (ii) from our PBEsol angle-averaged 
xc hole density through Eq.~(\ref{e2}). Overall, these calculations 
of $F^{PBEsol}_{xc}$ agree well with each other,
confirming the assumption made
at the beginning of this section;
 only for 
$r_s\geq 10$ (when the electron density is very small) 
and $s\approx 1.5$ 
is the error introduced by the second procedure significant \cite{note2}.
The analytic fit for $\mathcal{H}(s)$ 
used to construct our PBEsol exchange hole does not exactly 
reproduce the PBEsol enhancement factor, but the difference 
is small as shown in Fig.~\ref{f4}. At this point, we also 
note that the parametrization\cite{PBE,PRCVSCZB} of 
$H(r_s,\zeta,t)$ entering the analytic expression of 
$\epsilon^{PBEsol}_c$ reported in Ref.~\onlinecite{PRCVSCZB} 
does not reproduce exactly the real-space cutoff results, as shown in Figs.~7 and 8 of
Ref.~\onlinecite{PBW}.

\section{Wavevector analysis of the jellium xc surface energy}
\label{sec3}

The xc surface energy, $\sigma_{xc}$, can be 
defined as the xc energy cost per
per unit area to create a planar surface by cutting the bulk. 
In a jellium model, in which the 
electron system is translationally invariant in the plane of the surface, and
assuming the surface to be normal to the $z$-axis, 
the surface energy can be written as follows\cite{PCP}
\begin{equation}
\sigma_{xc}=\int^{\infty}_{0}d\left(k\over 2k_F\right)\,\gamma_{xc}(k),
\label{e15}
\end{equation}
where\cite{note1}
\begin{equation}
\gamma_{xc}(k)=2\,{k_F\over\pi}\int_{-\infty}^{+\infty}dz\,n(z)\,b_{xc}(k,z)
\label{e16}
\end{equation}
is the wavevector-resolved xc surface energy, and
\begin{equation}
b_{xc}(k,z)=4\pi\int_0^\infty du\,u^2\,
{\sin ku\over ku}\left[\bar n_{xc}(z,u)-\bar n^{unif}_{xc}(u)\right].
\label{e17}
\end{equation}
Equations (\ref{e15})-(\ref{e17}) comprise an angle-averaged three-dimensional
wavevector analysis of the surface xc energy into contributions from density fluctuations
of various wavevectors $k$, following from the Fourier transform of the Coulomb 
interaction in Eq. (\ref{e2}).
In these and subsequent equations,
$k_F = (3\pi^2 \bar n)^{1/3}$ is the bulk (not the local) Fermi wavevector, and $r_s$ 
is the bulk (not the local) density parameter.

The exact low-wavevector limit of $\sigma_{xc}$ is known to be\cite{LP1}
\begin{equation}
\gamma_{xc}(k\rightarrow 0)=\frac{k_F}{4\pi}(\omega_s-\frac{1}{2}\omega_p)k,
\label{e17b}
\end{equation} 
where $\omega_p=(4\pi\bar{n})^{1/2}$ and $\omega_s=\omega_p/\sqrt 2$ are the
bulk- and surface-plasmon energies, and $\bar{n}$ is the bulk density.
Eq.~(\ref{e17b}) was used in the wavevector-interpolation 
approach reported in Refs.~\onlinecite{LP1}, \onlinecite{CPT}, 
and \onlinecite{YPKFA}, and it was naturally recovered by 
the RPA approach reported in Ref.~\onlinecite{PCP}.

Taking into account that $\bf{k}\equiv(\bf{k_{||}},\bf{k_z})$, $\bf{k_{||}}$ 
being a wavevector parallel to the surface, Eq.~(\ref{e17}) can be expressed as
follows\cite{PCP}
\begin{eqnarray}
b_{xc}(k,z)&=&
\left[{1\over 2}\int_{-k}^{+k}{dk_{z}\over k}\int_{-\infty}^{+\infty}dz'\,
{\rm e}^{ik_z(z-z')}\right.\cr\cr
&\times&\left.\bar n_{xc}(k_{\parallel};z,z')-\bar n_{xc}^{unif}(k)\right].
\label{e18}
\end{eqnarray}
In the case of RPA and TDDFT calculations, we use the 
fluctuation-dissipation theorem\cite{HG,LP1,GL} to 
derive $n_{xc}(k_{\parallel};z,z')$ and
$\bar n_{xc}^{unif}(k)$ from the coupling-constant dependent density-response functions 
$\chi_\lambda^{unif}(k)$ and $\chi_\lambda(k_\parallel;z,z')$, as follows
\cite{PCP,PE}
\begin{equation}
\bar{n}_{xc}^{unif}(k)={1\over\bar n}
\left[-\frac{1}{\pi}\int_0^1d\lambda\int_0^\infty
d\omega\chi_\lambda^{unif}(k,i\omega)-\bar n\right]
\label{e19}
\end{equation}
and 
\begin{eqnarray}
\bar{n}_{xc}(k_{\parallel};z,z')&=&{1\over n(z)}\left[-\frac{1}{\pi}\int^{1}_{0}
d\lambda\int^{\infty}_{0}d\omega\right.\cr\cr
&\times&\left.\chi_\lambda(z,z';k_{\parallel},i\omega)-n(z)\delta(z-z')\right],
\label{e20}
\end{eqnarray}
$\chi_\lambda^{unif}(k,\omega)$ and $\chi_\lambda(z,z';k_{\parallel},i\omega)$ 
being 3D and 2D Fourier transforms of the corresponding density-response function
$\chi_{\lambda}({\bf r},{\bf r}';\omega)$. 
In the framework of TDDFT (our benchmark for this work), the density-response function
$\chi_{\lambda}({\bf r},{\bf r}';\omega)$ satisfies 
a Dyson-like equation of the form\cite{GDP}
\begin{eqnarray}
&&\chi_{\lambda}({\bf r},{\bf r}';\omega)=
\chi_0({\bf r},{\bf r}';\omega)+\int d\R_1\,d\R_2\,
\chi_0({\bf r},{\bf r}_1;\omega)\cr\cr
&\times& \left\{{\lambda\over |\R_1-\R_2|}+
f_{xc,\lambda}[n](\R_1,\R_2;\omega)\right\}\,
\chi_{\lambda}({\bf r}_2,{\bf r}';\omega),
\label{e21}
\end{eqnarray}
where $\chi_0({\bf r},{\bf r}';\omega)$ is the density-response function of
non-interacting KS electrons (which is exactly expressible in terms of KS
orbitals~\cite{GK}) and
$f_{xc,\lambda}[n]({\bf r},{\bf r}';\omega)$ is the {\it unknown}
$\lambda$-dependent dynamic xc kernel. When
$f_{xc,\lambda}[n]({\bf r},{\bf r}';\omega)$ is taken to be zero,
Eq.~(\ref{e21}) reduces to the RPA density-response function. 
If the interacting  density response function 
$\chi_{\lambda}({\bf r},{\bf r}';\omega)$ is replaced 
by the noninteracting KS density-response function 
$\chi_0({\bf r},{\bf r}';\omega)$, then Eqs. (\ref{e19}) 
and (\ref{e20}) yield their {\it exchange-only} counterparts.

In the calculations presented below, we have considered, as in
Ref.~\onlinecite{PCP}, a jellium slab of background thickness 
$a=2.23\lambda_F$ 
(where $\lambda_F = 2\pi/k_F$)
and bulk parameter $r_s=2.07$. This slab 
corresponds to about four atomic layers of Al(100).

For the GGA calculations of $\gamma_{xc}(k)$, the function 
$b_{xc}(k,z)$ entering Eq.~(\ref{e16}) is taken from Eq.~(\ref{e17}) 
with the xc-hole densities calculated as reported in 
(i) Ref.~\onlinecite{PW} for $\bar{n}_{xc}^{unif}(u)$, (ii)
Refs.~\onlinecite{PBW} and \onlinecite{EP} for 
$\bar{n}_{xc}^{PBE}(z,u)$, and (iii) Section II 
above for $\bar{n}_{xc}^{PBEsol}(z,u)$.

For the exact-exchange, exact-RPA, and TDDFT calculations of $\gamma_{xc}(k)$,
the function $b_{xc}(k,z)$ entering Eq.~(\ref{e16}) is taken from 
Eq.~(\ref{e18}) with the xc hole densities calculated from 
Eqs.~(\ref{e19}) and (\ref{e20}). In the case of the TDDFT calculations, 
we use the accurate static xc kernel reported and used in Ref.~\onlinecite{PP}. 
This kernel, which is based on a parametrization\cite{CSOP} of the 
diffusion Monte Carlo (DMC) calculations reported in Ref.~\onlinecite{MCS} 
for the uniform electron gas, 
was constructed for jellium surfaces where
neglect of the $\omega$-dependence does not introduce significant errors,
and is expected to yield exact results in the 
limits of small and large wavevectors. 
Our numerical scheme was described in detail in
Ref.~\onlinecite{PCP}, where only RPA calculations were reported.
%
\begin{figure}
\includegraphics[width=\columnwidth]{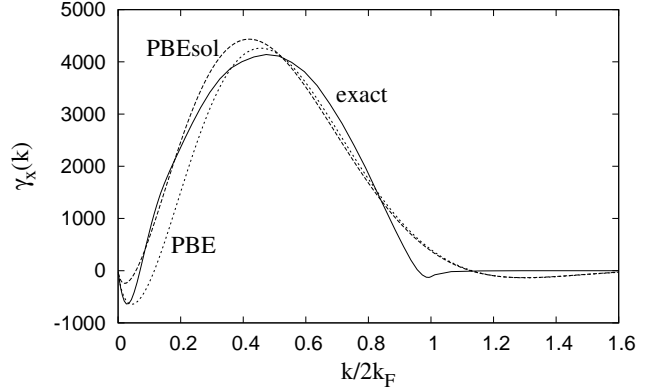}
\caption{PBE, PBEsol, and exact wavevector-resolved exchange surface energies
$\gamma_x(k)$, versus $k/2k_{F}$, for a jellium slab of thickness 
$a=2.23\lambda_F$ and $r_s=2.07$. The semilocal PBE and 
PBEsol calculations have been performed from
non-oscillatory parametrizations of the dimensionless exchange-hole shapes
$J^{PBE}$ and $J^{PBEsol}$, respectively. The area under each 
curve represents the corresponding exchange surface energy:
$\sigma^{PBE}_x=2164\; \mathrm{erg}/\mathrm{cm}^{2}$,
$\sigma^{PBEsol}_x=2424\; \mathrm{erg}/\mathrm{cm}^{2}$, and
$\sigma^{exact}_x=2348\; \mathrm{erg}/\mathrm{cm}^{2}$.}
\label{f5}
\end{figure}
%

\begin{figure}
\includegraphics[width=\columnwidth]{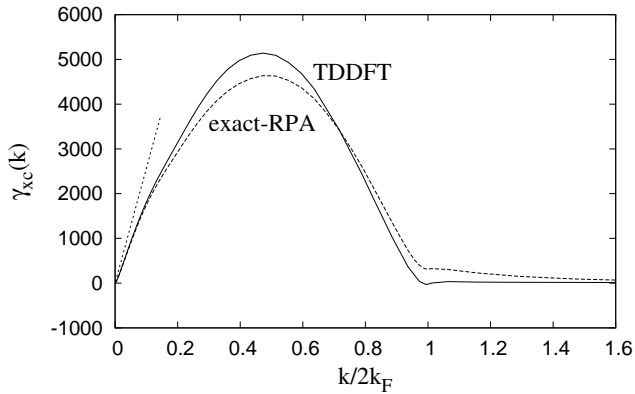}
\caption{Exact-RPA and benchmark TDDFT wavevector-resolved xc surface energies
$\gamma_{xc}(k)$, versus $k/2k_{F}$, for a jellium slab of 
thickness $a=2.23\lambda_F$ and $r_s=2.07$. The area under 
each curve represents the corresponding xc surface energy:
$\sigma^{RPA}_{xc}=3091\; \mathrm{erg}/\mathrm{cm}^{2}$ and
$\sigma^{TDDFT}_{xc}=3090\; \mathrm{erg}/\mathrm{cm}^{2}$. 
The straight dotted line represents the universal low-wavevector limit of
Eq.~(\ref{e17b}).}
\label{f6}
\end{figure}

\begin{figure}
\includegraphics[width=\columnwidth]{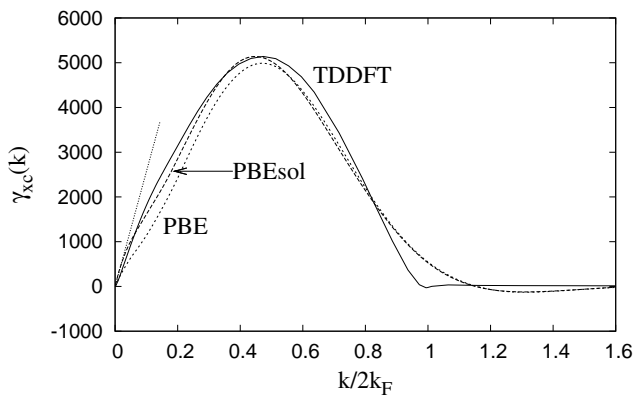}
\caption{PBE, PBEsol, and benchmark TDDFT wavevector-resolved xc surface energies
 $\gamma_{xc}(k)$, versus $k/2k_{F}$, for a jellium slab of thickness $a=2.23\lambda_F$ 
 and $r_s=2.07$. The semilocal PBE and PBEsol calculations have been performed from 
 non-oscillatory parametrizations of the dimensionless exchange-hole shapes $J^{PBE}$ and 
 $J^{PBEsol}$, respectively. The area under each curve represents the corresponding xc 
surface energy:
$\sigma^{PBE}_{xc}=2885\; \mathrm{erg}/\mathrm{cm}^{2}$,
$\sigma^{PBEsol}_{xc}=3027\; \mathrm{erg}/\mathrm{cm}^{2}$, and
 $\sigma^{TDDFT}_{xc}=3090\; \mathrm{erg}/\mathrm{cm}^{2}$. The straight dotted line 
represents the universal low-wavevector limit of Eq.~(\ref{e17b}).}
\label{f7}
\end{figure}

\begin{figure}
\includegraphics[width=\columnwidth]{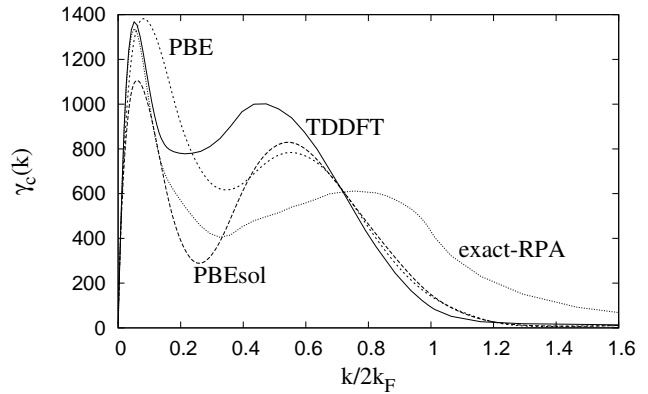}
\caption{PBE, PBEsol, exact-RPA, and benchmark TDDFT wavevector-resolved correlation 
 surface energies $\gamma_c(k)$, versus $k/2k_{F}$, for a jellium slab of thickness 
 $a=2.23\lambda_F$ and $r_s=2.07$. The area under each curve represents the corresponding 
correlation surface energy:
$\sigma^{PBE}_{c}=720\; \mathrm{erg}/\mathrm{cm}^{2}$,
$\sigma^{PBEsol}_{c}=604\; \mathrm{erg}/\mathrm{cm}^{2}$,
$\sigma^{RPA}_{c}=743\; \mathrm{erg}/\mathrm{cm}^{2}$, and
$\sigma^{TDDFT}_{c}=742\; \mathrm{erg}/\mathrm{cm}^{2}$.}
\label{f9}
\end{figure}

The wavevector-resolved exact-exchange surface energy 
$\gamma_{x}(k)$ is shown in Fig.~\ref{f5}. Figures~\ref{f6}-\ref{f7} 
and Fig.~\ref{f9} show, respectively, the wavevector-resolved xc 
and correlation-only surface energies $\gamma_{xc}(k)$ and
$\gamma_c(k)$. Figure~\ref{f5} shows that $\gamma_x^{PBEsol}(k)$ 
improves over PBE, as expected, and is close to the exact 
$\gamma_x(k)$ for intermediate values of the wavevector. 
At large values of the wavevector, both PBE and PBEsol correctly 
recover the nonoscillatory LSDA (see Figs.~1 and 2 of Ref.
\onlinecite{PCP}); differences between this nonoscillatory 
PBEsol (and also LSDA and PBE) and the exact $\gamma_x(k)$ at 
these large values of $k$ are due 
to inaccuracy of the nonoscillatory model employed
in Eq. (\ref{e10})
near $k=2k_F$.
%
%

In Fig.~\ref{f6}, we compare the wavevector-resolved 
exact-RPA surface energy (as reported in Ref.~\onlinecite{PCP}) 
with its TDDFT counterpart 
(which we have not reported elsewhere,
and which required three months of computation). 
In the long-wavelength limit ($k\to 0$), both RPA and TDDFT 
calculations approach the exact low-wavevector 
limit of Eq.~(\ref{e17b}). In the large-$k$ limit, 
where RPA is wrong, our TDDFT approach (which reproduces 
accurately the xc energy of the uniform 
electron gas) is expected to be accurate. Furthermore, the uniform-gas-based isotropic 
xc kernel that we use in our TDDFT calculation has been shown recently to yield 
essentially the same two-dimensional (2D) wavevector analysis as a more sophisticated 
high-level correlated approach (the inhomogeneous Singwi-Tosi-Land-Sj\"olander method) 
which does not use an isotropic kernel derived from the uniform gas.\cite{CPDGP} Hence, 
we take the TDDFT wavevector-resolved surface energy represented in Fig.~\ref{f6} by a 
solid line as the benchmark curve against which we compare various GGA's.

%
%

 Figure \ref{f7} shows our wavevector-resolved TDDFT surface energy together 
 with its PBE and PBEsol counterparts. $\gamma_{xc}^{PBEsol}$ 
is nearly exact at small wavevectors, where it matches the exact 
initial slope of Eq.~(\ref{e17b}). At intermediate wavevectors,
 $\gamma_{xc}^{PBEsol}$ is very close to our benchmark 
 TDDFT calculation. 
As in the case of wavevector-resolved exchange-only surface 
 energies, PBE and PBEsol calculations correctly recover the 
nonoscillatory 
 LSDA and differ from the more accurate TDDFT calculation due to
the inaccuracy of the nonoscillatory model
of the exchange-hole shape [see, e.g., 
 Eq.~(\ref{e10})]. 
Figure \ref{f7} shows that $\gamma_{xc}^{PBEsol}$ 
nicely
matches our benchmark
TDDFT calculation at low and intermediate wavevectors, so we 
recommend using the 
model PBEsol xc hole, in all solid-state calculations where
the hole is needed but full nonlocality is not important, instead of the more expensive TDDFT.
We recall that the 
system-averaged
hole, unlike the energy density, is uniquely defined, and is an observable at full 
coupling strength \cite{PBW}.
%

Figure \ref{f9} exhibits the wavevector-resolved correlation-only PBE, PBEsol, 
exact-RPA, and TDDFT surface energies. We observe that at long wavelengths
($k\to 0$), where the LSDA is known to fail badly, and at short wavelengths (large
 $k$), where RPA is wrong, the GGA's under consideration are considerably close to 
 our benchmark TDDFT calculations. At intermediate wavevectors, however, GGA's cannot 
 describe $\gamma_c$ accurately, although PBEsol has been shown in 
Fig.~\ref{f7}  to give a very good description of
 $\gamma_{xc}$. This is due to a cancellation of the errors introduced at these 
 wavevectors within the exchange and correlation contributions to $\gamma_{xc}$, which 
almost cancel each other \cite{note3}.  

\section{ Conclusions}
\label{sec4}

 We have constructed a PBEsol angle-averaged xc hole $\bar{n}_{xc}(\R,u)$ that satisfies 
 known exact constraints and recovers the recently reported PBEsol xc energy functional. 
 Our construction of the PBEsol xc hole 
begins from and appropriately modifies
the sharp cutoff
{\it correlation} 
 hole of Ref.~\onlinecite{PBW} and the smooth {\it exchange} hole of 
 Ref.~\onlinecite{EP}. We also generalize [see Eq.~(\ref{e7})] the sharp cutoff 
 procedure for the correlation hole to any GGA which has a positive gradient expansion 
 coefficient. 

 We have found that our PBEsol xc hole describes accurately the wavevector-resolved xc 
 jellium surface energy for all values of the wavevector, thus providing support for the 
PBEsol GGA for solids and surfaces.

\acknowledgments
L.A.C. and J.P.P. acknowledge NSF support (Grant No. DMR05-01588). J.M.P.
acknowledges partial support by the Spanish MEC (grant No. FIS2006-01343 and
CSD2006-53) and the EC 6th framework Network of Excellence NANOQUANTA. L.A.C.
thanks the Donostia International Physics Center (DIPC) where this work was started.


\newpage

\begin{thebibliography}{[Vo]}

\bibitem{KS}
{\ W. Kohn and L.J. Sham},
Phys. Rev. $\mathbf{140}$, A1133 (1965).
%
\bibitem{LP1}
{\ D.C. Langreth and J.P. Perdew},
Phys. Rev. B $\mathbf{15}$, 2884 (1977); $\mathbf{21}$, 5469 (1980);
$\mathbf{26}$, 2810 (1982).
%
\bibitem{PCP}
{\ J.M. Pitarke, L.A. Constantin, and J.P. Perdew},
Phys. Rev. B $\mathbf{74}$, 045121 (2006).
%
\bibitem{PBE}
{\ J.P. Perdew, K. Burke, and M. Ernzerhof},
Phys. Rev. Lett. $\mathbf{77}$, 3865 (1996).
%
\bibitem{PBW}
{\ J.P. Perdew, K. Burke and Y. Wang},
Phys. Rev. B $\mathbf{54}$, 16533 (1996); ibid. $\mathbf{57}$, 14999 (1998) (E).
%
\bibitem{EP}
{\ M. Ernzerhof and J.P. Perdew},
J. Chem. Phys. $\mathbf{109}$, 3313 (1998).
%
\bibitem{CPT}
{\ L.A. Constantin, J.P. Perdew, and J. Tao},
Phys. Rev. B $\mathbf{73}$, 205104 (2006).
%
\bibitem{TPSS}
{\ J. Tao, J.P. Perdew, V.N. Staroverov, and G.E. Scuseria},
Phys. Rev. Lett. $\mathbf{91}$, 146401 (2003).
%
\bibitem{PCSB}
{\ J.P. Perdew, L.A. Constantin, E. Sagvolden and K.Burke},
Phys. Rev. Lett. $\mathbf{97}$, 223002 (2006).
%
\bibitem{AK}
{\ P.R. Antoniewicz and L. Kleinman},
Phys. Rev. B $\mathbf{31}$, 6779 (1985).
%
%
\bibitem{MB}
{\ S.-K. Ma and K.A. Brueckner},
Phys. Rev. $\mathbf{165}$, 18 (1968).
%
\bibitem{PRCVSCZB}
{\ J.P. Perdew, A. Ruzsinszky, G.I. Csonka, O.A. Vydrov,
G.E. Scuseria, L.A. Constantin, X. Zhou, and  K. Burke},
Phys. Rev. Lett. {\bf 100}, 136406 (2008). Erratum submitted.
%
\bibitem{AM05}
{\ R. Armiento and A. E. Mattsson},
Phys. Rev. B $\mathbf{72}$, 085108 (2005).
%
\bibitem{KM}
{\ W. Kohn and A. E. Mattsson},
Phys. Rev. Lett. $\mathbf{81}$, 3487 (1998).
%
%
\bibitem{Furche1}
{\ M.P. Johansson, A. Lechtken, D. Schooss, M.M. Kappes, and F. Furche},
Phys. Rev. A $\mathbf{77}$, 053202 (2008) .
%
\bibitem{Gabor}
{\ G.I. Csonka, A. Ruzsinszky, J.P. Perdew, and S. Grimme}, 
J. Chem. Theory Comput. $\mathbf{4}$, 888 (2008).
%
\bibitem{WVK}
{\ R. Wahl, D. Vogtenhuber, and G. Kresse}, 
Phys. Rev. B $\mathbf{78}$, 104116 (2008).
%
\bibitem{PW1}
{\ J.P. Perdew and Y. Wang},
Phys. Rev. B $\mathbf{45}$, 13244 (1992).
%
\bibitem{OP}
{\ G.L. Oliver and J.P. Perdew},
Phys. Rev. A $\mathbf{20}$, 397 (1979).
%
\bibitem{PW}
{\ J.P. Perdew and Y. Wang},
Phys. Rev. B $\mathbf{46}$, 12947 (1992).
%
\bibitem{HJS}
{\ T.M. Henderson, B.G. Janesko, and G.E. Scuseria},
 J. Chem. Phys. $\mathbf{128}$, 194105 (2008).
%
\bibitem{PT}
{\ J.P. Perdew, J. Tao, V.N. Staroverov, and G.E. Scuseria},
J. Chem. Phys. $\mathbf{120}$, 6898 (2004).
%
\bibitem{note2}
We expect that our xc hole will have practically the same accuracy as the PBEsol xc energy
functional for the 
lattice-constant tests presented in Ref. \cite{PRCVSCZB}, because the PBEsol exchange is very
accurately fitted in Figs. 4 and 5.  It was shown in the Supporting Information of Ref. 
\cite{PRCVSCZB} that 
the lattice constants are not sensitive to the details of 
the gradient dependence of the correlation functional
(which is less well fitted in Figs. 4 and 5).
%
\bibitem{note1}
This equation holds for a semi-infinite jellium system. For a jellium slab, the 
right-hand side of Eq.~(\ref{e16}) should be divided by a factor of 2.
%
\bibitem{YPKFA}
{\ Z. Yan, J.P. Perdew, S. Kurth, C. Fiolhais, and L. Almeida},
Phys. Rev. B $\mathbf{61}$, 2595 (2000).
%
\bibitem{HG}
{\ J. Harris and A. Griffin},
Phys. Rev. B, $\mathbf{11}$, 3669 (1975).
%
\bibitem{GL}
{\ O. Gunnarsson and B.I. Lundqvist},
Phys. Rev. B $\mathbf{13}$, 4274 (1976).
%
\bibitem{PE}
J.M. Pitarke and A.G. Eguiluz, Phys. Rev. B $\mathbf{57}$, 6329 (1998);
$\mathbf{63}$, 045116 (2001).
%
\bibitem{GDP}
{\ E.K.U. Gross, J.F. Dobson and M. Petersilka}, in
\emph{Density Functional Theory II}, Vol.181 of
\emph{Topics in Current Chemistry}, edited by
R.F. Nalewajski (Springer, Berlin, 1996), p.81.
%
\bibitem{GK}
{\ E.K.U. Gross and W. Kohn},
Phys. Rev. Lett. $\mathbf{55}$, 2850 (1985).
%
\bibitem{PP}
J.M. Pitarke and J.P. Perdew,
Phys. Rev. B $\mathbf{67}$, 045101 (2003).
%
\bibitem{CSOP}
{\ M. Corradini, R. Del Sole, G. Onida and M. Palummo},
Phys. Rev. B $\mathbf{57}$, 14569 (1998).
%
\bibitem{MCS}
{\ S. Moroni, D.M. Ceperley, and G. Senatore},
Phys. Rev. Lett. $\mathbf{75}$, 689 (1995).
%
%
\bibitem{CPDGP}
L.A. Constantin, J. M. Pitarke, J. F. Dobson, A. Garc\'\i a-Lekue, and J.P. Perdew, 
Phys. Rev. Lett. {\bf 100}, 036401 (2008). 
%
%
%
%
%
%
\bibitem{note3}
Figs. 6 and 9 show that for small wavevectors ($k/2k_F<0.05$) and for
$k/2k_F$ between 0.25 and 0.4,
$\gamma_x^{PBE}(k)$ and $\gamma_c^{PBE}(k)$ are more accurate than their PBEsol analogs.
However, due to error cancellation between exchange and correlation,
$\gamma_{xc}^{PBEsol}(k)$ is considerably closer to 
$\gamma_{xc}^{TDDFT}(k)$ than is $\gamma_{xc}^{PBE}(k)$ for $k/2k_F\leq 0.5$ (see Fig. 8).




\end{thebibliography}
\end{document}